\PassOptionsToPackage{unicode}{hyperref}
\PassOptionsToPackage{hyphens}{url}
\documentclass[
]{article}
\usepackage{amsmath,amssymb}
\usepackage{setspace}
\usepackage{iftex}
\ifPDFTeX
  \usepackage[T1]{fontenc}
  \usepackage[utf8]{inputenc}
  \usepackage{textcomp} 
\else 
  \usepackage{unicode-math} 
  \defaultfontfeatures{Scale=MatchLowercase}
  \defaultfontfeatures[\rmfamily]{Ligatures=TeX,Scale=1}
\fi
\usepackage{lmodern}
\ifPDFTeX\else
\fi
\IfFileExists{upquote.sty}{\usepackage{upquote}}{}
\IfFileExists{microtype.sty}{
  \usepackage[]{microtype}
  \UseMicrotypeSet[protrusion]{basicmath} 
}{}
\makeatletter
\@ifundefined{KOMAClassName}{
  \IfFileExists{parskip.sty}{%
    \usepackage{parskip}
  }{
    \setlength{\parindent}{0pt}
    \setlength{\parskip}{6pt plus 2pt minus 1pt}}
}{
  \KOMAoptions{parskip=half}}
\makeatother
\usepackage{xcolor}
\usepackage[margin=1in]{geometry}
\usepackage{longtable,booktabs,array}
\usepackage{calc} 
\usepackage{etoolbox}
\makeatletter
\patchcmd\longtable{\par}{\if@noskipsec\mbox{}\fi\par}{}{}
\makeatother
\IfFileExists{footnotehyper.sty}{\usepackage{footnotehyper}}{\usepackage{footnote}}
\makesavenoteenv{longtable}
\usepackage{graphicx}
\makeatletter
\def\maxwidth{\ifdim\Gin@nat@width>\linewidth\linewidth\else\Gin@nat@width\fi}
\def\maxheight{\ifdim\Gin@nat@height>\textheight\textheight\else\Gin@nat@height\fi}
\makeatother
\setkeys{Gin}{width=\maxwidth,height=\maxheight,keepaspectratio}
\makeatletter
\def\fps@figure{htbp}
\makeatother
\NewDocumentCommand\citeproctext{}{}

\makeatletter
 \let\@cite@ofmt\@firstofone
 \def\@biblabel#1{}
 \def\@cite#1#2{{#1\if@tempswa , #2\fi}}
\makeatother
\newlength{\cslhangindent}
\newlength{\csllabelwidth}
\newenvironment{CSLReferences}[2] 
 {\begin{list}{}{%
  \setlength{\itemindent}{0pt}
  \setlength{\leftmargin}{0pt}
  \setlength{\parsep}{0pt}
  \ifodd #1
   \setlength{\leftmargin}{\cslhangindent}
   \setlength{\itemindent}{-1\cslhangindent}
  \fi
  \setlength{\itemsep}{#2\baselineskip}}}
 {\end{list}}
\usepackage{calc}

\usepackage[hang,flushmargin]{footmisc}
\usepackage{titling}
\usepackage{amsmath,amssymb}

\ifLuaTeX
  \usepackage{selnolig}  
\fi
\usepackage{bookmark}
\IfFileExists{xurl.sty}{\usepackage{xurl}}{} 
\hypersetup{
  pdftitle={How many federal employees are not satisfied? Using response times to estimate population proportions under the survey variable cause model.},
  hidelinks,
  pdfcreator={LaTeX via pandoc}}

\title{How many federal employees are not satisfied? Using response
times to estimate population proportions under the survey variable cause
model.}
\author{\newline\\
Jonathan Auerbach\\
Department of Statistics\\
George Mason University\\
\href{mailto:jauerba@gmu.edu}{\nolinkurl{jauerba@gmu.edu}}}
\date{}

\begin{document}
\maketitle
\begin{abstract}
\noindent We propose a statistical model to estimate population
proportions under the survey variable cause model (Groves 2006)---the
setting in which the characteristic measured by the survey has a direct
causal effect on survey participation. For example, we estimate employee
satisfaction from a survey in which the decision of an employee to
participate depends on their satisfaction. We model the time at which a
respondent `arrives' to take the survey, leveraging results from the
counting processes literature that has been developed to analyze similar
problems with survival data. Our approach is particularly useful for
nonresponse bias analysis because it relies on different assumptions
than traditional adjustments such as poststratification, which assumes
the common cause model---the setting in which external factors explain
the characteristic measured by the survey and participation. Our
motivation is the Federal Employee Viewpoint Survey, which asks federal
employees whether they are satisfied with their work organization. Our
model suggests that the sample proportion overestimates the proportion
of federal employees that are not satisfied with their work organization
even after adjustment by poststratification. Employees that are not
satisfied likely select into the survey, and this selection cannot be
explained by personal characteristics like race, gender, and occupation
or work-place characteristics like agency, unit, and location.
\end{abstract}

\setstretch{1.5}
\subsection{1. Introduction}\label{introduction}

We consider a survey administered to a population of \(N\) individuals.
The survey asks respondents to answer a yes or no question. We assume a
sample of \(n\) individuals complete the survey, \(n < N\). However, the
sample is biased because employees that would answer no are more likely
to complete the survey. The goal is to estimate the proportion of nos in
the population of \(N\) individuals from the biased sample of \(n\)
responses. We depart from traditional survey analysis in that we do not
adjust the survey data using auxiliary variables such as age, gender,
and occupation. Instead, we assume event-time paradata is available with
each response---paradata such as the time at which the respondent
provided their response---and we leverage tools from the counting
processes literature designed to model event-time data.

Our work is organized in three sections. In the remainder of this
section, we present our motivating example, the Federal Employee
Viewpoint Survey, which asks federal employees whether they are
satisfied with their work organization. We show how poststratification
can fail or even increase bias if the characteristic measured
(satisfaction) has a direct causal effect on survey participation or the
auxiliary variables used for adjustment. In Section 2, we outline the
model proposed to estimate the population proportion using response
times. Our model is derived using the partial likelihood approach of
David R. Cox (1972), but inference is different because the population
proportion is unknown (whereas in survival analysis the number ``at
risk'' is typically known). In Section 3, we discuss extensions and
limitations. Additional details are included in the Appendix.

\subsubsection{1.1 How many federal employees are not
satisfied?}\label{how-many-federal-employees-are-not-satisfied}

Our departure from traditional practice is motivated by the Federal
Employee Viewpoint Survey (FEVS), an annual survey of workplace climate
within the federal government. The survey is administered to more than
1.7 million federal employees each year. We limit our interest to one
item on the survey, which asks respondents ``Considering everything, how
satisfied are you with your organization?'' Respondents provide their
answer to this question on a five-point Likert scale, and we consider a
federal employee to be ``satisfied'' if they responded ``4 - satisfied''
or ``5 - very satisfied''. The remaining respondents who answered the
question are considered ``not satisfied.''

We wish to estimate the percentage of federal employees that are not
satisfied, and we examine the data collected in 2022. Identical
examinations for the years 2018 and 2023 yield similar findings. Note
that not all federal employees are eligible to complete the survey. When
we refer to all federal employees, we mean all eligible federal
employees.

In 2022, 31 percent of federal employees responding to the FEVS
indicated they were not satisfied with their organization. The problem
with this sample statistic is that participation in the survey is
voluntary. Only a third of eligible federal employees responded to the
survey in 2022. The low response rate suggests the potential for
nonresponse bias. (Nonresponse here refers to unit nonresponse. Item
nonresponse was less than five percent.) Indeed, it would be hard to
believe that employee satisfaction plays no role in an employee's
decision whether to participate in the survey.

Federal guidelines state that a nonresponse bias analysis should be
conducted if the expected unit response rate is below 80 percent. The
FEVS methodology adjusts for unit nonresponse by poststratification
using an extensive list of employee characteristics observed for all
federal employees, including agency, subagency, supervisory status, sex,
minority status, age group, tenure, full-time or part-time status, and
location. Poststratification is standard for nonresponse bias
adjustment, see Lohr (2021, chap. 8), Heeringa et al. (2017, sec. 2.7),
and Groves et al. (2011, chap. 6). The details are documented in
Appendix F of the FEVS methodology and summarized in the appendix of
this paper.

The details are not particularly relevant to our discussion. What is
relevant is the underlying assumption. Consider the following
hypothetical example, chosen for simplicity (the numbers do not reflect
the actual data). Suppose supervisors make up half the population but 90
percent of respondents. Further suppose that 80 percent of supervisors
are not satisfied, and 20 percent of non-supervisors. Then 74 percent of
the sample would report they are not satisfied when in reality only 50
percent of the population is not satisfied.

Under the assumption that supervisors who participate in the survey
represent all supervisors (and non-supervisors who participate represent
all non-supervisors), poststratification works by estimating the
population proportion separately for supervisors and non-supervisors (80
percent and 20 percent). The weighted average of these estimates, using
the known proportion of supervisors in the population (50 percent or
1/2) yields the population proportion: (80 + 20) / 2 = 50 percent.

Surprisingly, adjustment by poststratification does not materially
change the proportion estimated from the FEVS data. After adjustment, an
estimated 34 percent of all federal employees were not satisfied with
their work organization in 2022 with a standard error of essentially 0.
The question is whether selection persists by some mechanism not
accounted for by the aforementioned characteristics. Or worse, whether
poststratification introduces more bias than it removes, as we now
discuss.

\subsubsection{1.2 Causal models for
nonresponse}\label{causal-models-for-nonresponse}

Groves et al. (2011) identifies five causal models for nonresponse,
consistent with the MCAR/MAR/NMAR taxonomy for missing data mechanisms
outlined by Little and Rubin (2019). We assume employee satisfaction is
measured without error, and we limit our attention to the first three
models that do not include measurement error: the separate causes model,
the common causes model, and the survey variable cause model. Graphical
representations of these models are reproduced from Groves et al. (2011)
in Figure 1 below. \(Y\) represents the characteristic being measured,
\(P\) represents the decision to participate (the response propensity),
and \(X\) and \(Z\) are external characteristics that determine \(Y\)
and \(P\). Thick arrows represent a direct causal relationship while
thin arrows represent an indirect statistical relationship induced by
direct dependence on a common external factor.

\begin{figure}
\centering
\includegraphics{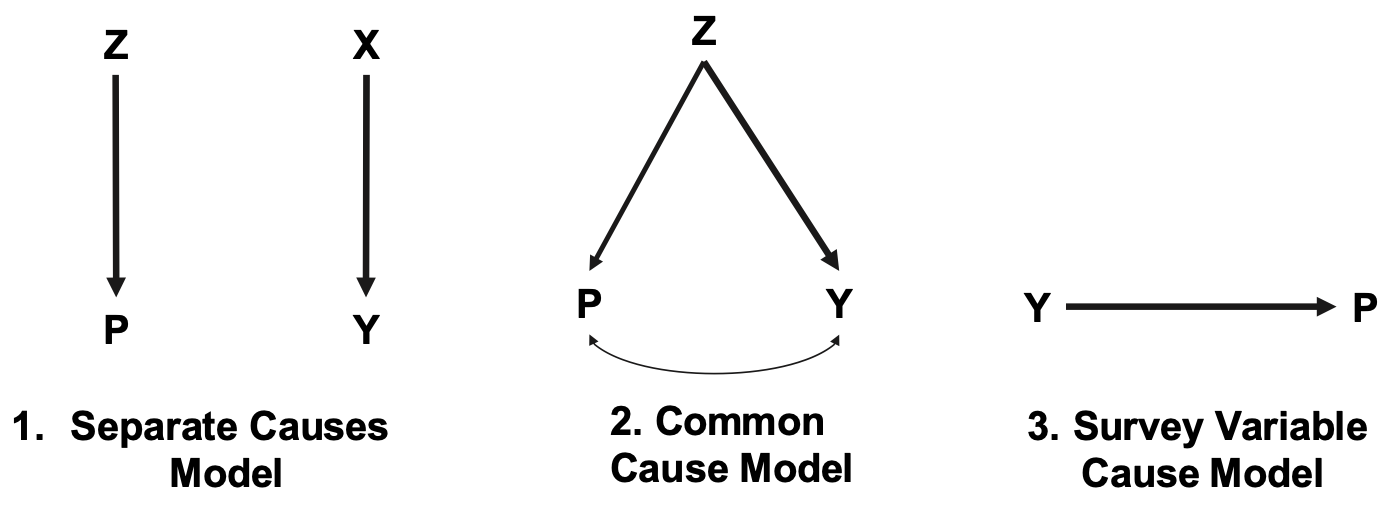}
\caption{This figure shows three of five causal models for nonresponse,
reproduced from Groves et al. (2011). Estimating population proportions
with sample proportions is justified under the separate causes model.
Poststratification is justified under the common cause model. Under the
survey cause model, however, poststratification can fail or even
introduce additional bias. In such cases, modeling the arrival of
respondents in a response time analysis can reveal the bias and provide
a better estimate.}
\end{figure}

Under the separate causes model, federal employees do not select into
the employee satisfaction survey because satisfaction is uncorrelated
with the factors that determine selection. (There is no arrow between
\(Z\) and \(X\) or between \(Z\) and \(Y\) in Figure 1.) This
corresponds with employees missing completely at random (MCAR) and
suggests that the 31 percent of federal employees that indicated they
were not satisfied reflects the population of all federal employees.

Under the common cause model, federal employees select into the survey
according to external characteristics that also determine whether an
employee is satisfied. For example, employees with supervisor status may
be less satisfied and more likely to participate in the survey. This
corresponds with employees missing at random (MAR)---in particular that,
among federal employees with the same characteristics, employees are
missing completely at random. (The thin line between P and Y is induced
by the variable Z.) FEVS methodology adjusts for an extensive list of
employee characteristics, suggesting that if the separate cause model
holds, the adjusted proportion of 34 percent is accurate or at least
adjusts the sample proportion in the right direction.

The final possibility is the survey variable cause model. Satisfaction
determines whether an employee chooses to participate in the survey.
This corresponds with employees not missing at random (NMAR). Satisfied
employees are fundamentally different than not satisfied employees in
their motivation to participate in the survey, even after accounting for
observed characteristics. In this case, both the 31 and 34 percent
estimates are potentially incorrect. Moreover, if satisfaction
determines the auxiliary characteristics such as supervisor status,
full-time status, or location, poststratification can introduce
bias---an example of ``collider bias'' that often arises in causal
inference.

Consider the following change to the hypothetical example from Section
1.1 in which the decision to respond to the survey is now a proxy for
engagement. Suppose satisfaction and engagement are largely unrelated,
but satisfaction and engagement are both important factors in attaining
supervisory status. In this example, the sample proportion would not
have much bias because satisfied and not satisfied employees are
similarly engaged and thus participate in the survey at similar rates.
Poststratification introduces bias if, for example, not satisfied
supervisors are more engaged than their satisfied supervisor
counterparts while not satisfied non-supervisors are less engaged.

To make this example concrete, consider the following hypothetical data
in Table 1, chosen again for simplicity---the numbers do not reflect the
actual data. Suppose half the population is not satisfied and
approximately a quarter are engaged and participate in surveys. The
proportion of not satisfied respondents is 45 + 2 = 47 percent, which is
close to the true 50 percent. However, when broken down by supervisory
status, 50 percent of supervisor respondents are not satisfied and 20
percent of non-supervisor respondents. Since half the population are
supervisors, the adjusted estimate by poststratification is (50 + 20) /
2 = 35 percent, a bias five times higher than the sample proportion.
\vspace{1em}

\newpage

Table 1. Hypothetical example in which poststratification increases bias
\vspace{-0.75em}

\begin{longtable}[]{@{}crr@{}}
\toprule\noalign{}
& Population (\%) & Respondents (\%) \\
\midrule\noalign{}
\endhead
\bottomrule\noalign{}
\endlastfoot
\textbf{Supervisor} & & \\
Satisfied & 40 & 45 \\
Not satisfied & 10 & 45 \\
\textbf{Non-Supervisor} & & \\
Satisfied & 10 & 8 \\
Not satisfied & 40 & 2 \\
\textbf{Totals} & 100 & 100 \\
\end{longtable}

This example represents one of many possible ways in which inference is
complicated under the survey variable cause model. An alternative
approach to poststratification is to examine responses over time in a
response time analysis. A response time analysis can identify and
correct for bias, as we now demonstrate.

\subsubsection{1.3 Response time analysis}\label{response-time-analysis}

We examine the timing of the responses from both satisfied and not
satisfied federal employees in a response time analysis. Our response
time analysis suggests that contrary to the poststratification analysis,
satisfaction has a large effect on whether an employee chooses to
participate in the survey. Our analysis follows common practice to
monitor daily response rates. See Kreuter (2013, chap. 2) for a
discussion of how paradata like response times are used to identify and
adjust for nonresponse bias. We deviate from current practice by
modeling the hazard ratio, which measures the rate at which not
satisfied employees select into the survey relative to satisfied
employees.

Our analysis is summarized in two figures, Figure 2 and Figure 3, in
which we separate respondents by the day they responded. We start with
Figure 2. The x-axis denotes the day of each response over the six-week
response period. Each point represents the proportion of respondents
that indicated they were not satisfied among those that responded that
day. Lines represent two standard errors. As mentioned before, less than
5 percent of respondents did not answer the question of interest. These
are excluded from the figure. The proportion of respondents that are not
satisfied is indicated by the solid line labeled ``sample proportion.''

\begin{figure}
\centering
\includegraphics{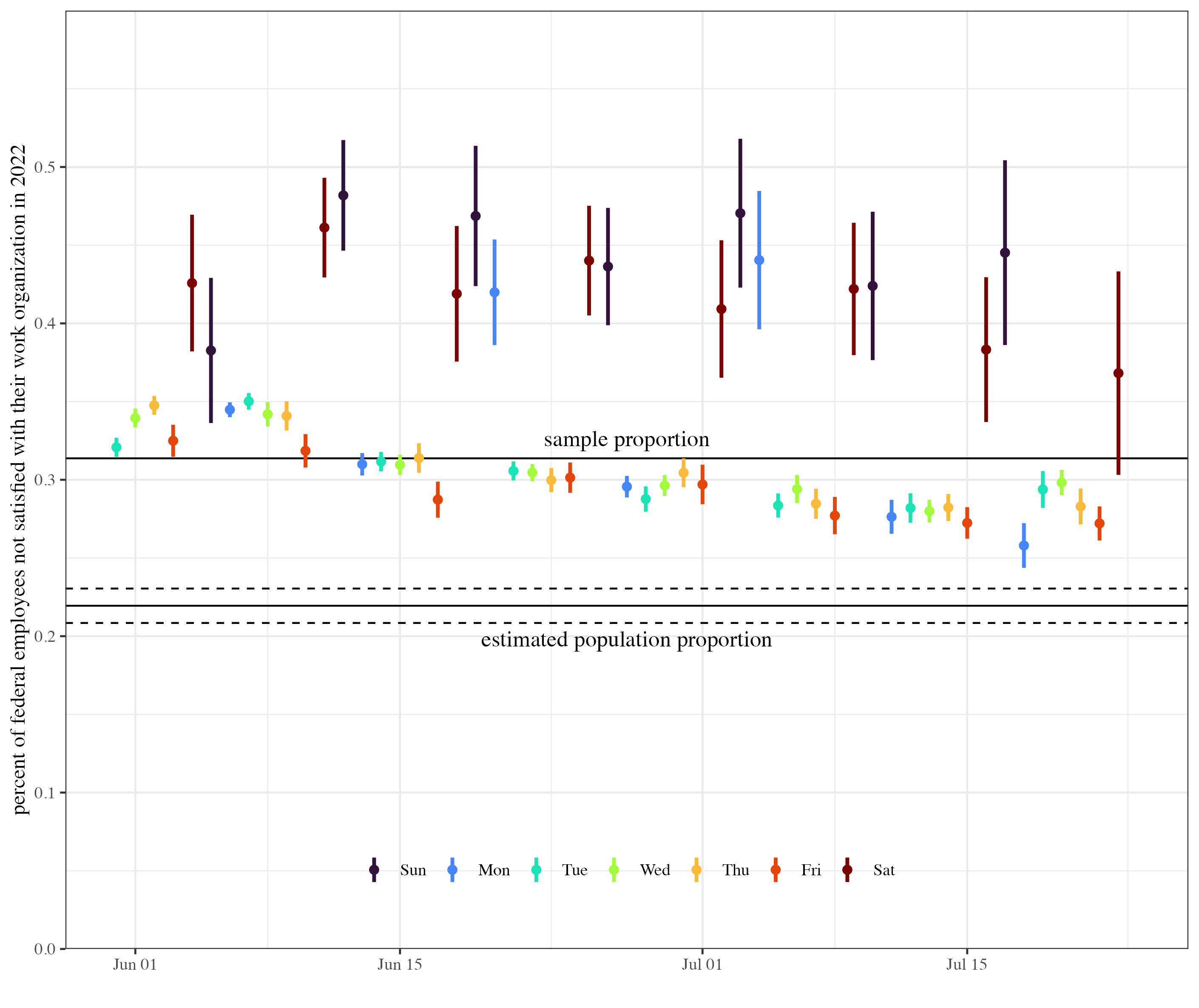}
\caption{This figure shows the proportion of federal employees that
indicated they were not satisfied with their work organization in the
2022 Federal Employee Viewpoint Survey. The proportion (points) is
calculated each day the survey was open and colored by day of the week.
Lines represent two standard errors. Respondents provide their answer to
the question ``Considering everything, how satisfied are you with your
organization?'' on a five-point Likert scale, and we consider a federal
employee to be ``satisfied'' if they responded ``4 - satisfied'' or ``5
- very satisfied''. The remaining respondents who answered the question
are considered ``not satisfied.'' The gradual decline of the proportion
of not satisfied employees suggests not satisfied employees select into
the survey and thus the sample proportion (top solid line) overestimates
the actual proportion (as estimated by bottom solid line, dotted lines
represent two standard errors)}
\end{figure}

Figure 2 reveals two patterns. We highlight these patterns by coloring
the points and lines by the day of the week. Weekend days are colored
black (Sunday) and brown (Saturday), while weekdays are colored using
``rainbow colors'' blue (Monday) to red (Friday). The first pattern is
that a greater proportion of individuals that respond on the weekend and
federal holidays (Monday June 20th and July 4th) are not satisfied,
although the large standard errors reflect the fact that less than 2
percent of responses to the survey occur on the weekend or a holiday. We
ignore these respondents for the moment.

The second pattern is the trend among the remaining respondents, in
which the percent that are not satisfied declines over time. When the
survey opened on May 31st, slightly under 35 percent of respondents
indicated they were not satisfied. By the close of the survey, slightly
under 27 percent indicated they were not satisfied. The overall sample
proportion is approximately the midpoint (35 + 27) / 2 = 31 percent.

The trend is statistically significant. Thus, instead of assuming the
satisfaction rate is the same for respondents and nonrespondents, as is
assumed in poststratification, one might extrapolate the trend for a
naive estimate of the population proportion. For a back-of-the-envelope
calculation, note that it took approximately six weeks for a third of
the sample to respond, over which time the sample proportion declined
from 35 percent to 27 percent, a drop of 8 percentage points. If that
trend continued, the remaining two-thirds of the population might have
responded over the following twelve weeks and shown a further drop of 16
percentage points. In this case, roughly 11 percent of respondents would
indicate they were not satisfied in the final few days, and the
percentage of all federal employees that are not satisfied would be
approximately (35 + 11) / 2 = 23 percent.

Extrapolation is not ideal because we can only speculate how respondents
might arrive were the survey open longer. While we believe the trend
would likely continue, in theory the percent of not satisfied
respondents could stay at 27 percent or the trend could even reverse.
Fortunately, while extrapolation is illuminating, it is not necessary.
As long as respondents are capable in theory of completing the survey at
any time during the response period and the response probabilities
change slowly over time, the declining participation rate in Figure 2
identifies the population proportion.

We provide the technical details of this argument in Section 2. We
conclude this section with a visual argument aimed at sidestepping the
technical details and providing additional intuition. Before introducing
this argument, however, note that the back-of-the-envelope extrapolation
agrees with the proposed approach, which is indicated by the solid line
labeled ``estimated population proportion.'' The dotted lines denote two
standard errors. The estimate suggests that employees that are not
satisfied likely select into the survey, and this selection cannot be
explained by personal characteristics like race, gender, and occupation
or work-place characteristics like agency, unit, and location.

Figure 3 provides intuition behind the proposed approach. Let \(\pi\)
denote the population proportion of federal employees that are not
satisfied with their work organization. If we knew \(\pi\), we could
calculate the daily response rate separately for the satisfied and not
satisfied employees. (The daily response rate or hazard is the number of
employees that responded on a given day divided by the number of
employees that had not yet responded prior to that day.) The daily
response ratio (hazard ratio) is the ratio of the daily response rates.

\begin{figure}
\centering
\includegraphics{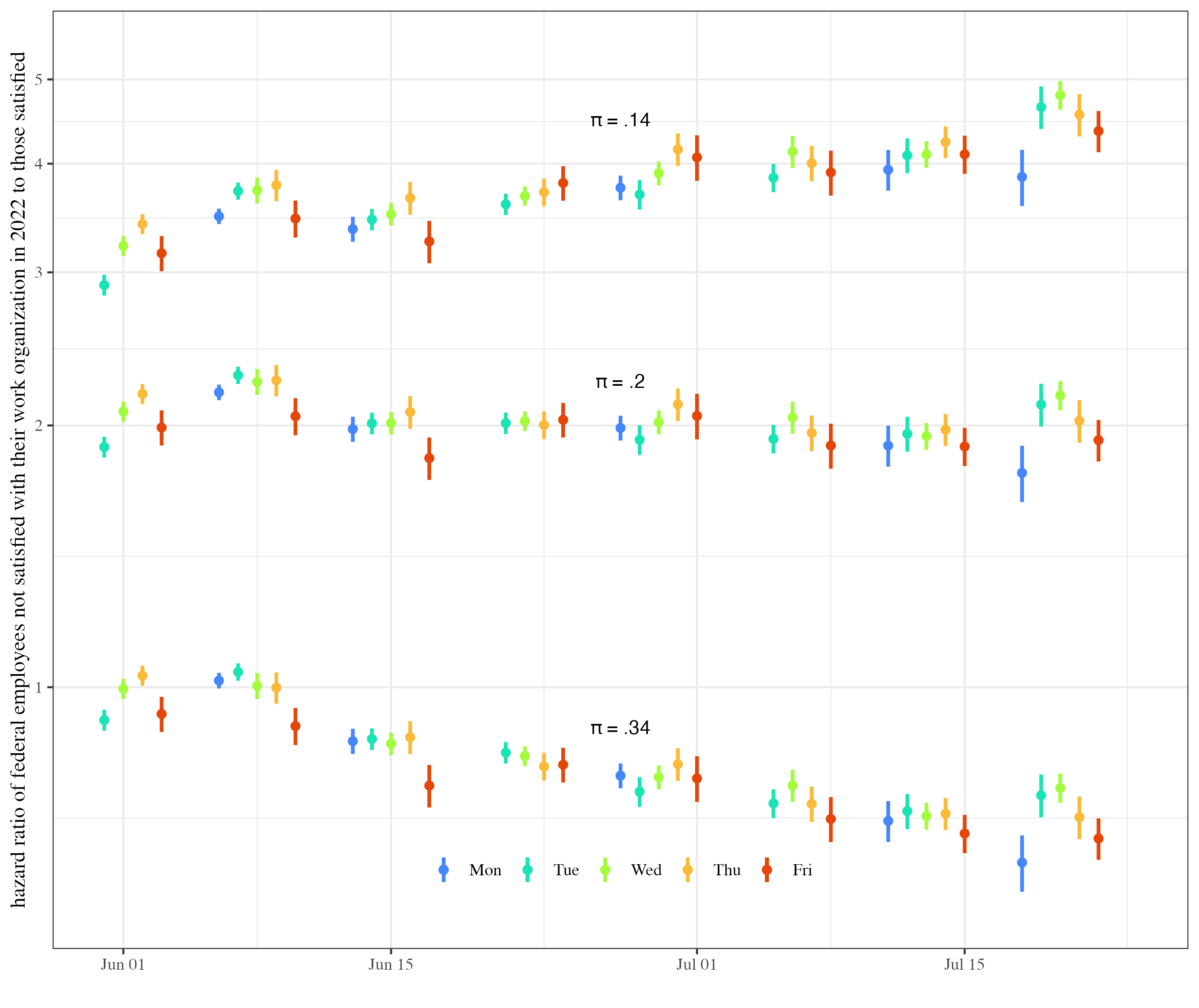}
\caption{This figure shows the hazard ratio (points) of not satisfied to
satisfied federal employees each day, assuming not satisfied employees
make up \(\pi\) percent of the population. The hazard ratio is the ratio
of the daily response rates: A ratio of 2 means a response was twice as
likely to come from a not satisfied employee as a satisfied employee.
The hazard ratio is shown for all weekdays, excluding holidays, and is
colored by day of the week. Lines represent two standard errors. The
y-axis is show on the log scale. If the population proportion
\(\pi = .2\), then the hazard is constant at around 2 (not satisfied
employees are twice as likely than satisfied employees to complete the
survey on any day), and day-to-day variation is largely explained by
chance. If \(\pi = .34\) or \(\pi = .14\), the hazard ratio varies
considerably over time and day-to-day or week-to-week variation cannot
be explained by chance.}
\end{figure}

Figure 3 shows the daily response ratio (hazard ratio) for \(\pi\) equal
to .14, .2, and .34. As in Figure 2, the days are colored by the day of
the week, although here weekends and holidays are excluded. Lines
represent two standard errors. We find that the ratio for \(\pi = .2\)
stays around 2, and day to day variation is largely explained by chance.
In contrast, the ratio for \(\pi = .14\) and \(\pi = .34\) change
systematically over time, and day-to-day or week-to-week variation
cannot easily be explained by chance.

Response time analysis thus yields the following insight. If
\(\pi \approx .2\), not satisfied employees are twice as likely to
participate as satisfied employees at any time period. The steady
decline of the sample proportion in Figure 2 is a statistical artifact,
reflecting the fact that as not satisfied employees select into the
survey, there are fewer not satisfied employees left to take the survey.
If \(\pi \approx .34\), as suggested by the sample proportion and
poststratification, the daily response rate fluctuates over time.
Initially satisfied and not satisfied employees are equally likely to
respond, while over time satisfied employees become nearly twice as
likely to respond.

We believe \(\pi \approx .2\) is the simplest explanation for the data
unless a mechanism can be identified which explains the declining daily
response ratio (hazard ratio). Note that such an explanation would have
to explain the fact that these results replicate across agencies and in
the years 2018 and 2023. We think the gradually declining ratio in
Figure 3 for \(\pi = .34\) is an artifact from overestimating the number
of not satisfied employees.

Our technical development in Section 2 follows this observation, that if
the response ratio (hazard ratio) does not change too abruptly---that
is, it is smooth or otherwise well approximated by a piecewise constant
function---\(\pi\) is identified and estimable. The proportional hazard
model (David R. Cox 1972) commonly used in survival analysis naturally
arises, a connection we now make precise. The main difference between
our setting and survival analysis is that in our setting the group
proportions, \(\pi\), are unknown and the hazard ratio is a nuisance
parameter. In contrast, population parameters such as \(\pi\) are known
in the typical survival setting, and the hazard ratio represents the
treatment effect, the parameter of interest.

\subsection{2. Methodology}\label{methodology}

We consider a survey administered to a finite population of \(N\)
individuals. Each individual belongs to one of two groups. Let \(X_i\)
be the binary variable indicating the group assignment of the \(i\)th
individual, \[
  X_i =
  \begin{cases}
    1,&\text{if individual $i$ belongs to the group of interest},\\
    0,&\text{if individual $i$ belongs to the reference group.}
  \end{cases}
\]

The goal is to estimate the proportion of individuals belonging to the
group of interest, \(\pi = \sum_{i=1}^N X_i \, / \, N\). Our running
example is the climate survey described in Section 1, which is sent to
approximately 1.7 million federal employees each year. The parameter of
interest is the percentage of federal employees that are not satisfied
with their work organization.

The problem is that the survey is closed before all individuals respond.
Let \(T_i > 0\) denote the time individual \(i\) would hypothetically
respond to the survey were it not closed. We assume the survey is closed
at a predetermined time \(\tau\) after which \(T_i\) is right censored
and the corresponding label \(X_i\) is unobserved. That is, we assume
the tuple \((T_i, \, X_i)\) is observed for all individuals for which
\(T_i < \tau\). Among the remaining individuals for which
\(T_i \geq \tau\), \(X_i\) is unobserved, and it is only known that
\(T_i \geq \tau\).

The challenge is to estimate \(\pi\) from the data
\(\{ (T_i, \, X_i)\}_{T_i < \tau }\) and \(N\). In our motivating
example, 33 percent of eligible federal employees responded to the
survey during the six week response period in 2022. Of those who
responded, 31 percent reported that they were not satisfied with their
work organization. However, if dissatisfied employees are more likely to
respond before time \(\tau\), the proportion of dissatisfied employees
in the sample overestimates the population proportion \(\pi\), the
proportion of all dissatisfied employees.

\subsubsection{2.1 Model}\label{model}

We assume two sources of randomness. The first source of randomness
comes from the response times, \(T_i\), which we assume follow a
group-specific hazard that varies arbitrarily across time \[
\lambda_{X_i}(t)
=\lim_{h\downarrow0}
\frac{\Pr\bigl(t\le T_i<t+h\mid T_i>t,\,X_i\bigr)}{h}
=\frac{f_{X_i}(t)}{1 - F_{X_i}(t)},
\] where \(f_j(t)\) is the density of group \(j\in\{0,1\}\), and
\(F_j(t)\) is the cumulative distribution function. The hazard is the
instantaneous probability an individual responds at time \(t\) given
they have not yet responded. We assume \(\lambda_j\) is uniformly
bounded away from 0 and infinity so that each individual has a finite
response time. We refer to this condition as positivity.

Recall that the labels \(X_i\) are unobserved when \(T_i > \tau\). We
refer to this missingness as unit nonresponse because the individual did
not respond before the deadline \(\tau\). Unit nonresponse is driven
entirely by the randomness of \(T_i\).

In addition, to unit nonresponse, we assume a second source of
randomness due to missingness at random before \(\tau\), in which
\(T_i\) is observed but not \(X_i\). We refer to this missingness as
item nonresponse because the individual responded before the deadline
\(\tau\) but did not complete the item. Both sources of missingness are
tracked by the binary random variable \[
\delta_i =
\begin{cases}
1, & T_i\le\tau\text{ and the label }X_i\text{ is recorded},\\
0, & \text{otherwise (either }T_i>\tau\text{ or }X_i\text{ is missing).}
\end{cases}
\] We assume item nonresponse is uninformative, \[
\delta_i \;\perp\!\!\!\perp\;(T_i,X_i)
\;\bigm|\;
T_i<\tau.
\]

Recall that in our motivating example, 33 percent of eligible federal
employees responded to the survey during the six week response period in
2022. Of those that responded, 95 percent answered the question
``Considering everything, how satisfied are you with your
organization?'' Thus, the unit response rate in 2022 was 33 percent and
the item response rate in 2022 was 95 percent.

Note that we deviate from the survival literature in that we assume the
response time is always observed provided it occurs before time
\(\tau\). Only the group labels are missing. In contrast, the survival
literature typically assumes both the response time and the group label
are missing. See Kalbfleisch and Prentice (2002, chap. 3).

Also note that in some applications it also makes sense to treat the
group labels, \(X_i\), as random, see for example Tsiatis (2006). We do
not consider \(X_i\) random. Note however that the group labels, when
ordered by the random event times, are random, as discussed in Section
2.2.

\subsubsection{2.2 Estimation}\label{estimation}

We follow David R. Cox (1972) and construct the likelihood by
conditioning on the arrival times. See also David R. Cox and Oakes
(1984). Let \(T_{(i)}\) denote the \(i\)th ordered event times \[
0 < T_{(1)} < \cdots < T_{(n)} < \tau,
\] where \(n\) is the (random) sample size
\(n = \sum_{i=1}^N 1\{T_i < \tau \}\). We assume no ties. Also let
\(X_{[i]}\) and \(\delta_{[i]}\) denote the ``concomitant'' group label
and nonresponse indicator, corresponding with \(T_{(i)}\). That is, the
sets \(\{X_{[i]}\}_{i=1}^n\) and \(\{\delta_{[i]}\}_{i=1}^n\) denote the
sets \(\{X_{i}\}_{i=1}^n\) and \(\{\delta_{i}\}_{i=1}^n\) after ordering
by \(T_{i}\).

The conditional probability of label \(X_{[i]} = j\) at time
\(T_{(i)} = t\) given the events proceeding \(t\),
\(\mathcal{F}_{T_{(i-1)}}\), is \[
\Pr\bigl(X_{[i]}=j\mid T_{(i)}=t,\, \mathcal{F}_{T_{(i-1)}}\bigr)
= \frac{\lambda_j(t) \, N_j(t)}
       {\lambda_0(t)\,N_0(t) + \lambda_1(t) \, N_1(t)},
\quad j\in\{0,1\}
\] where \(N_0(t) = N \, (1-\pi) - \sum_{T_{(i)} < t} (1 - X_{[i]})\) is
the number from the reference group who have no yet responded before
time \(t\) and are ``at risk,'' while
\(N_1(t) = N \, \pi - \sum_{T_{(i)} < t} X_{[i]}\) is the number from
the group of interest.

Cox's partial likelihood is the product of these conditional
probabilities. It corresponds to the conditional likelihood in which the
event times, \(T_{(i)}\), are fixed and the concomitant labels
\(X_{(i)}\) are random. There are several equivalent representations \[
\begin{aligned}
\mathcal L \left (\pi, \lambda_0(.), \lambda_1(.) \right )
&= \prod_{i:\,\delta_{[i]}=1}
    \Pr\bigl(X_{[i]} \mid T_{(i)},\, \mathcal F_{T_{(i-1)}}\bigr) \\[1em]
&= \prod_{i:\,\delta_{[i]}=1}
  \frac{\lambda_{X_{[i]}}\bigl(T_{(i)}\bigr)\,N_{X_{[i]}}\bigl(T_{(i)}\bigr)}
       {\lambda_{0}\bigl(T_{(i)}\bigr)\,N_{0}\bigl(T_{(i)}\bigr)
        \;+\;\lambda_{1}\bigl(T_{(i)}\bigr)\,N_{1}\bigl(T_{(i)}\bigr)} \\[2em]
&= \prod_{i:\,\delta_{[i]}=1}
  \frac{ \left( \lambda_{0}(T_{(i)})\,N_{0}(T_{(i)}) \right)^{1-X_{[i]}}
        \left ( \lambda_{1}(T_{(i)})\,N_{1}(T_{(i)}) \right)^{X_{[i]}}}
       {\lambda_{0}(T_{(i)})\,N_{0}(T_{(i)})
        \;+\;\lambda_{1}(T_{(i)})\,N_{1}(T_{(i)})} \\[2em]
&= \prod_{i:\,\delta_{[i]}=1}
  \frac{\rho(T_{(i)})^{\,X_{[i]}}\;N_1\bigl(T_{(i)}\bigr)^{X_{[i]}}\;
      N_0\bigl(T_{(i)}\bigr)^{1-X_{[i]}}}
     {N_0\bigl(T_{(i)}\bigr)+\rho(T_{(i)})\,N_1\bigl(T_{(i)}\bigr)},
\end{aligned}
\] where \[
\rho(t)=\frac{\lambda_{1}(t)}{\lambda_{0}(t)}
\] is the (possibly time‐varying) hazard ratio between the group of
interest and the reference group. Note that individuals with
\(\delta_{[i] = 0}\) contribute \(1\) to the likelihood because under
independence, marginalizing over the values of \(X_{[i]} = j\),
\(j \in \{0,1\}\) results in a conditional probability of \(1\).

The rationale of the partial likelihood is that the group-specific
hazard functions, \(\lambda_{j}(t)\), which determine the response
times, may be challenging to estimate in practice. Any number of factors
might determine whether an individual has the time to complete a survey
at any instant. However, the vast majority of those factors are shared
across groups, such that the hazard ratio is likely to be low
dimensional or even constant. So while it might be unreasonable to
assume a not satisfied employee has a constant probability of responding
at each instant over a given day, it is perhaps more reasonable to
assume that the probability a not satisfied employee responds at any
instant is proportional to a satisfied individual.

Identification of \(\pi\) requires that \(\rho(t)\) is approximately
constant on a nontrivial subset of \([0, \tau]\). We assume the hazard
ratio \(\rho(t)\) is piecewise constant. We partition the time axis into
\(K\) intervals \(\{I_k\}_{k=1}^K\), and assume \(\rho(t)=\rho_k\) when
\(t\in I_k\). The partial likelihood is then \[
\mathcal L (\pi,\rho, \ldots, \rho_K )
=\prod_{i:\,\delta_{[i]}=1}
\frac{\rho_{\,k[i]}^{\,X_{[i]}}\;N_1\bigl(T_{(i)}\bigr)^{X_{[i]}}\;
      N_0\bigl(T_{(i)}\bigr)^{1-X_{[i]}}}
     {N_0\bigl(T_{(i)}\bigr)+\rho_{\,k[i]}\,N_1\bigl(T_{(i)}\bigr)}.
\] where \(\rho_{\,k[i]}\) refers to the ``concomitant'' proportional
hazard, corresponding with the \(i\)th arrival at time \(T_{(i)}\).

Estimation of \(\pi\) and \(\rho_1, \ldots, \rho_K\) is done by
maximizing the partial likelihood. Note that \(N_0(t)\) and \(N_1(t)\),
the number of individuals that have yet to respond from each group at
time \(t\) (that is, the ``number at risk'' for response), are
unobserved even when the parameters are known due to item nonresponse.
We use consistent estimators
\[ \hat N_0(t) = \left (N - \sum_{T_{(i)} < t} (1 - \delta_{[i]}) \right ) \, (1-\hat \pi) - \sum_{T_{(i)} < t} (1 - X_{[i]}) \, \delta_{[i]}\]
and
\[\hat N_1(t) = \left (N - \sum_{T_{(i)} < t} (1 - \delta_{[i]}) \right )  \, \hat \pi - \sum_{T_{(i)} < t} X_{[i]} \, \delta_{[i]}.\]

\subsubsection{2.3 Standard Errors}\label{standard-errors}

The partial likelihood in Section 3.2 is a valid likelihood and thus the
usual asymptotic theory for maximum likelihood estimation holds with
some modifications to account for the fact that the score and Hessian
are the sums of dependent random variables. We follow the insight of
Aalen (1975, 1978) and subsequent development by Per Kragh Andersen and
Gill (1982) and Borgan (1984), which recognize that the score is a
martingale and well approximated by a normal distribution under the
martingale central limit theorem. See Per K. Andersen et al. (2012,
chap. II.5) and Fleming and Harrington (2013, chap. 5) for details.

Note that the proposed model is not a multiplicative intensity model
because \(N_0(t)\) and \(N_1(t)\) depend on the unknown parameter
\(\pi\). The violation is largely superficial since these functions are
still predictable. Nevertheless, we verify the conditions outlined by
Pul (1992), which generalizes the conditions stated in Borgan (1984).
Specifically, we proceed treating the partial likelihood as the full
likelihood and note conditions G1-G2 and L1-L3 of Pul (1992) are
satisfied. As \(N \rightarrow \infty\) and
\(n \, / \, N \rightarrow p\), \(0 < p < 1\), the maximizer of the
partial likelihood is approximately normal with variance equal to the
inverse of the Fisher information.

The log--partial‐likelihood is \[
\ell(\pi,\rho_1, \ldots, \rho_K )
=\sum_{i:\,\delta_{[i]}=1}
\Bigl[
  X_{[i]}\log\rho_{\,k[i]}
  + X_{[i]}\log N_1\bigl(T_{(i)}\bigr)
  + (1-X_{[i]})\log N_0\bigl(T_{(i)}\bigr)
  -\log\bigl(N_0\bigl(T_{(i)}\bigr)+\rho_{\,k[i]}N_1\bigl(T_{(i)}\bigr)\bigr)
\Bigr].
\] The score has entries \[
U_\pi = \frac{\partial}{\partial\pi} \, \ell(\pi,\rho_1, \ldots, \rho_K )
= \sum_{i:\,\delta_{[i]}=1}
\Biggl[
  X_{[i]}\,\frac{N}{N_1(T_{(i)})}
  \;-\;(1-X_{[i]})\,\frac{N}{N_0(T_{(i)})}
  \;-\;\frac{-N+\rho_{\,k[i]}\,N}{N_0(T_{(i)})+\rho_{\,k[i]}\,N_1(T_{(i)})}
\Biggr].
\] and \[
U_{\rho_k} = \frac{\partial}{\partial\rho_k} \, \ell(\pi,\rho_1, \ldots, \rho_K )
= \sum_{\substack{i:\,\delta_{[i]}=1\\k[i]=k}}
\Biggl[
  \frac{X_{[i]}}{\rho_k}
  \;-\;
  \frac{N_1\bigl(T_{(i)}\bigr)}
       {N_0\bigl(T_{(i)}\bigr)+\rho_k\,N_1\bigl(T_{(i)}\bigr)}
\Biggr].
\]

The Hessian has entries \[
H_{\pi\pi} = \frac{\partial^2}{\partial\pi^2} \, \ell(\pi,\rho_1, \ldots, \rho_K )
=
\sum_{i:\,\delta_{[i]}=1}
\Biggl[
  -\,\frac{N^2\,X_{[i]}}{N_1(T_{(i)})^2}
  \;+\;\frac{N^2\,(1-X_{[i]})}{N_0(T_{(i)})^2}
  \;-\;\frac{N^2\,(1-\rho_{\,k[i]})^2}{\left( N_0(T_{(i)}) + \rho_{\,k[i]}\,N_1(T_{(i)}) \right)^2}
\Biggr],
\]

\[
H_{\pi\rho_k} = \frac{\partial^2}{\partial\pi\,\partial\rho_k} \, \ell(\pi,\rho_1, \ldots, \rho_K )
=\sum_{\substack{i:\,\delta_{[i]}=1\\k[i]=k}}
-\;\frac{N\bigl[N_0(T_{(i)})+N_1(T_{(i)})\bigr]}{\left( N_0(T_{(i)}) + \rho_{\,k[i]}\,N_1(T_{(i)}) \right)^2},
\]

\[
H_{\rho_k \rho_k}  = 
\frac{\partial^2}{\partial\rho_k^2} \, \ell(\pi,\rho_1, \ldots, \rho_K )
=
\sum_{\substack{i:\,\delta_{[i]}=1\\k[i]=k}}
\Biggl[
  -\,\frac{X_{[i]}}{\rho_k^2}
  \;+\;\frac{N_1(T_{(i)})^2}{\left( N_0(T_{(i)}) + \rho_{\,k[i]}\,N_1(T_{(i)}) \right)^2}
\Biggr],
\] and for \(j\neq k\), \[
H_{\rho_j \rho_k} = \frac{\partial^2}{\partial\rho_j\,\partial\rho_k} \, \ell(\pi,\rho_1, \ldots, \rho_K ) = 0.
\]

The information for \(\pi\) is
\(I(\pi) = H_{\pi\pi} - \sum_{k=1}^K H^2_{\pi\rho_k} \, / \, H_{\rho_k \rho_k}\)
and the variance of \(\hat \pi\), the maximizer of the partial
likelihood, is approximately \(I(\hat \pi)^{-1}\).

\subsubsection{2.4 Application to the Federal Employee Viewpoint
Survey}\label{application-to-the-federal-employee-viewpoint-survey}

We apply the estimation strategy outlined in Section 2.2 to the data
described in Section 1 and the Appendix. Standard errors are calculated
as described in Section 2.3. The only part of the model we have direct
influence over is the piecewise constant function \(\rho(t)\). We tried
several specifications for \(\rho(t)\), and we estimated a similar
population proportion \(\pi\) for each one.

For example, we 1. fixed the hazard ratio as constant in \(t\),
\(\rho(t) = \rho\); 2. allowed the hazard ratio to vary by month and day
of the week; and 3. we allowed the hazard ratio to vary every
\(k = 10, 15, 20\), etc. week days (weekends and holidays were given a
separate hazard ratio). These approaches suggest the population
proportion \(\pi\) is approximately 22 percent with a standard error of
approximately 1 percent.

The model started to break down with \(k < 5\), for example when each
day or week received its own hazard ratio. This produced unrealistically
extreme values of \(\pi\) as well as extreme standard errors. This
behavior is explained by the fact that \(\pi\) is weakly identified if
\(\rho\) is not sufficiently smooth. Some of the behavior could also be
explained by within-day and within-week artifacts. For example,
respondents appear to prefer to complete the survey during work hours
and earlier in the week. These artifacts are avoided by assuming a
constant \(\rho\) at the day or week level.

\subsection{3. Discussion}\label{discussion}

We conclude from the response time analysis that approximately 20
percent of federal employees were not satisfied with their work
organization in 2022. At any instant during the response period, not
satisfied federal employees were twice as likely to respond to the
survey as satisfied federal employees. Furthermore, this selection
cannot be explained by auxiliary variables such as age, gender, and
occupation. A similar analysis performed in 2018 and 2023 produced
similar results.

The response time analysis suggests that the sample proportion
overestimates the true proportion by about fifty percent in 2022.
Moreover, poststratification adjusts the estimate in the wrong
direction, evidence that the common cause model used to justify
poststratification does not hold. Instead, poststratification introduces
a form of collider bias: the auxiliary variables used in
poststratification are likely themselves the result of employee
satisfaction. Federal employees may select employee characteristics like
supervisory status, occupation, or work unit based on their
satisfaction.

These findings assume the assumptions underlying the response time
analysis hold. There are several limitations to our analysis that could
be addressed in future work. We list four major areas of development
below.

We assume there are no other sources of total survey error in addition
to nonresponse bias such as measurement error and coverage error. For
example, the answers of respondents might not actually reflect their
satisfaction. Respondents may not read the question correctly or they
may be motivated to produce a false answer. The Federal Employee
Viewpoint Survey has over a hundred items. These could be used to adjust
for measurement error, possibly using an approach that combines the
counting process approach outlined here with a latent class approach,
for example, as outlined by Biemer (2011).

Another limitation is that we only consider a single binary variable.
The approach is easily extended to a continuous variable or multiple
variables with additional modeling assumptions. For example, responses
to the satisfaction item we examined is reported on the Likert scale.
Our analysis could easily be altered to treat that scale as categorical,
ordinal, or even continuous. For example, if satisfaction is a
continuous, normally distributed random variable, the responses on the
Likert scale might be treated as satisfaction percentiles. Instead of
estimating the sample proportion, we could estimate the mean and
variance of the underlying normal distribution.

A third limitation is that we did not include the covariates used for
poststratification. Covariates could be included by allowing the
baseline hazard or hazard ratio to vary by group. We did not include
covariates because we were worried about collider bias as discussed in
Section 1. Variation in the baseline hazard could also be modeled
without covariates, for example, by using a subject-specific random
effect (frailty term). Unless the distribution of the random effect is
highly skewed, however, the estimated proportion is unlikely to change
due to the massive size of the Federal Employee Viewpoint Survey.

A final limitation is that we assumed a piecewise constant hazard ratio.
We could have also estimated the ratio nonparametrically using, for
example, a spline or kernel. The piecewise constant model made sense in
our case because time naturally binned into days and weeks, although
these time units could easily be incorporated into a spline or kernel
approach. In addition, we could have applied these methods directly to
the full likelihood by assuming the hazards are continuous. However, the
standard errors using the partial likelihood are already small, so the
efficiency of the full likelihood is likely not worth the potential for
introducing bias through model misspecification.

\subsection{\texorpdfstring{References
\vspace{1mm}}{References }}\label{references}

\phantomsection\label{refs}

\subsection{Appendix}\label{appendix}

\subsubsection{A.1. The Federal Employee Viewpoint
Survey}\label{a.1.-the-federal-employee-viewpoint-survey}

The Federal Employee Viewpoint Survey (FEVS) is an annual climate survey
administered to approximately 1.7 million federal employees each year.
FEVS satisfies a federal law requiring each agency to conduct an annual
survey of its employees to assess a) leadership and management practices
that contribute to agency performance and b) employee satisfaction
(Public Law 108-136 Sec. 1128, codified 5 USC Sec. 7101). Federal
regulations specify an Annual Employee Survey of at least 16 items (5
CFR Part 250 Subpart C). The FEVS items vary each year, but always
include the Annual Employee Survey items, supplemented by a large number
of additional items. For example, the 2022 FEVS includes more than 122
items: 20 demographic items and 104 workplace climate items, as well as
additional items included at the request of individual agencies. Topics
include the pandemic, paid parental leave, and diversity, equity,
inclusion, and accessibility.

The responses are used to assess agency policies and management
practices, provide insights into performance, and identify actions to
improve workplace effectiveness (U.S. Office of Personnel Management
2022a). The survey is voluntary, however, and missing or incomplete
responses may provide a misleading picture on which to base these
assessments as discussed in Section 1. In such cases, the climate
reported by respondents may fail to reflect the climate experienced by
all eligible employees---a phenomenon known as nonresponse bias.

The response rate to FEVS, defined as the number of eligible employees
returning completed surveys divided by the number of eligible employees,
is displayed in Table A1 below. (The rate is calculated using the OPM
FEVS formula, see U.S. Office of Personnel Management 2022b.)
\vspace{1em}

Table A.1. FEVS Response Rates from 2013 to 2023 \vspace{-0.75em}

\begin{longtable}[]{@{}
  >{\raggedright\arraybackslash}p{(\columnwidth - 22\tabcolsep) * \real{0.1791}}
  >{\raggedright\arraybackslash}p{(\columnwidth - 22\tabcolsep) * \real{0.0746}}
  >{\raggedright\arraybackslash}p{(\columnwidth - 22\tabcolsep) * \real{0.0746}}
  >{\raggedright\arraybackslash}p{(\columnwidth - 22\tabcolsep) * \real{0.0746}}
  >{\raggedright\arraybackslash}p{(\columnwidth - 22\tabcolsep) * \real{0.0746}}
  >{\raggedright\arraybackslash}p{(\columnwidth - 22\tabcolsep) * \real{0.0746}}
  >{\raggedright\arraybackslash}p{(\columnwidth - 22\tabcolsep) * \real{0.0746}}
  >{\raggedright\arraybackslash}p{(\columnwidth - 22\tabcolsep) * \real{0.0746}}
  >{\raggedright\arraybackslash}p{(\columnwidth - 22\tabcolsep) * \real{0.0746}}
  >{\raggedright\arraybackslash}p{(\columnwidth - 22\tabcolsep) * \real{0.0746}}
  >{\raggedright\arraybackslash}p{(\columnwidth - 22\tabcolsep) * \real{0.0746}}
  >{\raggedright\arraybackslash}p{(\columnwidth - 22\tabcolsep) * \real{0.0746}}@{}}
\toprule\noalign{}
\begin{minipage}[b]{\linewidth}\raggedright
Year
\end{minipage} & \begin{minipage}[b]{\linewidth}\raggedright
2013
\end{minipage} & \begin{minipage}[b]{\linewidth}\raggedright
2014
\end{minipage} & \begin{minipage}[b]{\linewidth}\raggedright
2015
\end{minipage} & \begin{minipage}[b]{\linewidth}\raggedright
2016
\end{minipage} & \begin{minipage}[b]{\linewidth}\raggedright
2017
\end{minipage} & \begin{minipage}[b]{\linewidth}\raggedright
2018
\end{minipage} & \begin{minipage}[b]{\linewidth}\raggedright
2019
\end{minipage} & \begin{minipage}[b]{\linewidth}\raggedright
2020
\end{minipage} & \begin{minipage}[b]{\linewidth}\raggedright
2021
\end{minipage} & \begin{minipage}[b]{\linewidth}\raggedright
2022
\end{minipage} & \begin{minipage}[b]{\linewidth}\raggedright
2023
\end{minipage} \\
\midrule\noalign{}
\endhead
\bottomrule\noalign{}
\endlastfoot
Response (\%) & 48.2 & 46.8 & 49.7 & 45.8 & 45.5 & 40.6 & 42.6 & 44.3 &
33.8 & 35.3 & 38.9 \\
\end{longtable}

The FEVS methodology considers a survey response to be complete if the
respondent answers at least 25\% of the non-demographic items. Thus, the
response rates in Table 1 reflect a combination of unit and item
nonresponse---although in this report, unit nonresponse includes
employees who return incomplete surveys to remain consistent with FEVS
methodology. Information is available for all eligible respondents
(respondents and nonrespondents) from the sampling frame, a database
maintained by the U.S. Office of Personnel Management called the
Statistical Data Mart of the Enterprise Human Resources Integration
(EHRI-SDM). The information is used to adjust for nonresponse and
includes: subagency, supervisory status, sex, minority status, age
group, tenure as a federal employee, full-time or part-time status, and
location.

The FEVS methodology adjusts for nonresponse using sampling weights
equal to the inverse of the estimated probability (propensity) that an
employee is both eligible to respond and provides a complete response.
The weights are constructed as follows. First, eligible employees are
divided into groups (strata) according to whether they have similar
covariates in the sampling frame (i.e., grouped according to sex,
minority status, age, etc.) Second, the number of strata are reduced
using CHAID, a tree-like algorithm in which strata are iteratively
merged if the probability of a complete response is deemed the same
according to a chi-square test for independence (Kass 1980). Finally,
employees within strata are assumed to have the same probability of
providing a complete response, and the strata-specific weights (inverse
estimated probabilities) are raked so that weighted totals agreed with
the number of eligible employees by sex, subagency, and minority status
within agency. See Appendix F of U.S. Office of Personnel Management
(2022b) for details. No additional adjustment is performed for item
nonresponse.

In some years, only a random sample of eligible employees are invited to
complete the survey, and the inverse of the sampling probabilities serve
as base weights. (i.e., Instead of counting the number of employees
within each strata, the base-weighted number of employees is used.) In
addition, due to software limitations, a different algorithm was used to
reduce the number of strata prior to 2016.

\subsubsection{A.2 References}\label{a.2-references}

Kass, Gordon. 1980. An Exploratory Technique for Investigating Large
Quantities of Categorical Data. Journal of the Royal Statistical
Society, Series C. 29 (2), 119--127.
\url{https://doi.org/10.2307/2986296}

\emph{Title 5 U.S.C.} § 250 Subpart C. 2023.
\url{https://www.ecfr.gov/current/title-5/chapter-I/subchapter-B/part-250/subpart-C}

U.S. Office of Management and Budget. 2006. Standards and guidelines for
statistical surveys.
\url{https://www.whitehouse.gov/wp-content/uploads/2021/04/standards_stat_surveys.pdf}

U.S. Office of Personnel Management. 2022a. Federal Employee Viewpoint
Survey Results: Governmentwide Management Report.
\url{https://www.opm.gov/fevs/reports/governmentwide-reports/governmentwide-reports/governmentwide-management-report/2022/2022-governmentwide-management-report.pdf}

U.S. Office of Personnel Management. 2022b. Federal Employee Viewpoint
Survey Results: Technical Report.
\url{https://www.opm.gov/fevs/reports/technical-reports/technical-report/technical-report/2022/2022-technical-report.pdf}

\end{document}